\documentclass[prl,reprint,superscriptaddress]{revtex4-1}

\usepackage{graphicx}
\usepackage{xcolor}
\usepackage{dcolumn}
\usepackage{bm}
\usepackage{amssymb}
\usepackage{amsmath}
\usepackage{verbatim}
\usepackage[utf8]{inputenc}
\usepackage{tikz,hyperref}
\usepackage{dcolumn}

\definecolor{lime}{HTML}{A6CE39}
\DeclareRobustCommand{\orcidicon}{%
	\begin{tikzpicture}
	\draw[lime, fill=lime] (0,0) 
	circle [radius=0.16] 
	node[white] {{\fontfamily{qag}\selectfont \tiny ID}};
	\draw[white, fill=white] (-0.0625,0.095) 
	circle [radius=0.007];
	\end{tikzpicture}
	\hspace{-2mm}
}

\foreach \x in {A, ..., Z}{%
	\expandafter\xdef\csname orcid\x\endcsname{\noexpand\href{https://orcid.org/\csname orcidauthor\x\endcsname}{\noexpand\orcidicon}}
}

\newcommand{\vect}[1]{\boldsymbol{#1}}

\begin{document}

\title{Self-Interacting Dark Matter and Small-Scale Gravitational Lenses in Galaxy Clusters}

\author{Daneng Yang \orcidA{}}
\email{yangdn@mail.tsinghua.edu.cn}
\affiliation{Department of Physics, Tsinghua University, Beijing 100084, China}
\affiliation{Department of Physics and Astronomy, University of California, Riverside, California 92521, USA}

\author{Hai-Bo Yu \orcidB{}}
\email{haiboyu@ucr.edu}
\affiliation{Department of Physics and Astronomy, University of California, Riverside, California 92521, USA}

\date{\today}

\begin{abstract}
Recently, Meneghetti et al. reported an excess of small-scale gravitational lenses in galaxy clusters. We study its implications for self-interacting dark matter (SIDM), compared with standard cold dark matter (CDM). We design controlled N-body simulations that incorporate observational constraints. The presence of early-type galaxies in cluster substructures can deepen gravitational potential and reduce tidal mass loss. Both scenarios require a relatively high baryon concentration in the substructure to accommodate the lensing measurements, and their tangential caustics are similar. The SIDM substructure can experience gravothermal collapse and produce a steeper density profile than its CDM counterpart, leading to a larger radial galaxy-galaxy strong lensing cross section, although this effect is hard to observe. Our results indicate SIDM can provide a unified explanation to small-scale lenses in galaxy clusters and stellar motions in dwarf galaxies. 
\end{abstract}

\maketitle

{\noindent\bf Introduction.} Strong gravitational lensing is characterized by the existence of giant arcs, rings, and multiple images caused by the deflection of lights by massive foreground galaxies, groups, or galaxy clusters~\cite{1992grle.book.....S,Kneib:2012ip}. It provides a powerful tool for testing cosmological models~\cite{Cao:2011bg, Merten:2014wna}, determining the mass distribution of clusters~\cite{Newman:2012nv,Annunziatella:2017svn,Bonamigo:2018ioa}, probing substructures~\cite{Mao:1997ek,Dalal:2001fq,Keeton:2002qt,Xu:2009ch,Grillo:2014ofa,Natarajan:2017sbo,Nierenberg:2017vlg,Gilman:2019bdm,Minor:2020bmp} and dark matter properties~\cite{Vegetti:2014wza,Minor:2020hic,Gilman:2019nap,Andrade:2020lqq,He:2020rkj,Nadler:2021dft}. Recently, Meneghetti et al. reported that observed substructures in galaxy clusters are more efficient lenses than those predicted in simulations of standard cold dark matter (CDM)~\cite{Meneghetti:2020yif}, indicating that the former are more dense and compact. Other studies also show strong lensing clusters contain more substructures with high maximum circular velocities than predicted in CDM simulations~\cite{Grillo:2014ofa,Munari:2016uqd}. The tension could be resolved in CDM with a different implementation of baryonic feedback, resulting in dense substructures~\cite{Bahe:2021bcs,Robertson:2021}. 

We study self-interacting dark matter (SIDM)~\cite{Spergel:1999mh,Kaplinghat:2015aga,Tulin:2017ara} in light of small-scale gravitational lenses observed in~\cite{Meneghetti:2020yif}. In this scenario, dark matter collisions thermalize the inner halo over cosmological timescales~\cite{Dave:2000ar,Vogelsberger:2012ku,Rocha:2012jg,Zavala:2012us}. Compared with CDM, SIDM better explains diverse dark matter distributions inferred in a wide range of galactic systems from ultra-diffuse galaxies to galaxy clusters~\cite{Firmani:2000ce,Kaplinghat:2015aga,Kamada:2016euw,Creasey:2016jaq,Salucci:2018hqu,Ren:2018jpt,Kaplinghat:2019svz,Kaplinghat:2019dhn,Yang:2020iya,Sagunski:2020spe}. It may also explain the origin of supermassive black holes~\cite{Balberg:2001qg,Pollack:2014rja,Choquette:2018lvq,Feng:2020kxv}. The observations of dense substructures in galaxy clusters~\cite{Meneghetti:2020yif} seem to challenge SIDM, as the thermalization could lead to shallow density cores, which are prone to tidal disruption. In contrast to this general expectation, we will show in this work that the SIDM substructure can be as efficient as its CDM counterpart in producing small-scale lenses, after taking into account realistic stellar distributions and tidal environments.

We design N-body SIDM and CDM simulations to model the MACS J1206.2-0847 (MACSJ1206) cluster, one of the examples studied in~\cite{Meneghetti:2020yif}, and construct four benchmark cases, covering a representative range of halo concentration. The presence of early-type galaxies in the substructures could significantly accelerate the onset of SIDM gravothermal collapse~\cite{Balberg:2002ue,Koda:2011yb,Essig:2018pzq,Huo:2019yhk}. For all benchmarks, our simulated SIDM subhalos experience collapse after $6~{\rm Gyr}$ of tidal evolution and become more dense than their CDM counterparts, assuming a self-scattering cross section per mass of $\sigma/m=1~{\rm cm^2/g}$, which is relatively conservative~\cite{Tulin:2017ara}. For both SIDM and CDM scenarios, the simulated substructures require a high baryon concentration to be consistent with the observations in~\cite{Meneghetti:2020yif}.

We further model strong lensing observables and compute galaxy-galaxy strong lensing (GGSL) cross sections for the benchmarks. The tangential caustic predicted in the SIDM substructure is comparable to its CDM counterpart, while the former has a much larger radial GGSL cross section, with details depending on the source redshift and initial halo concentration. We will also show mock lensing images and discuss observational implications of a radial GGSL cross section.

{\noindent\bf Modeling the cluster system.} We model the host cluster using a static spherical potential characterized by a Navarro-Frenk-White (NFW) density profile~\cite{Navarro:1996gj} and fix its corresponding scale density and radius as $\rho_s=1.82 \times10^{6}~{\rm M_\odot/kpc^3}$ and $r_s=442~{\rm kpc}$, respectively. It well reproduces the projected total mass profile of MACSJ1206~\cite{2019A&A...631A.130B,2017A&A...607A..93C,2018ApJ...864...98B}. We set the apocenter of our simulated substructures to be $400~{\rm kpc}$, with a tangential velocity of $1000~{\rm km/s}$. During the tidal evolution, their distance to the host center oscillates in the range $\sim140\textup{--}{370}~{\rm kpc}$. The orbit controls the significance of tidal stripping, and hence the mass loss. In our setup, the mass of the simulated substructures at $6~{\rm Gyr}$ is ${\cal O}(10^{10})~{\rm M_\odot}$. The strong lensing analysis in~\cite{Meneghetti:2020yif} focuses on substructures within $15\%$ of the virial radius of the host cluster, which is $300~{\rm kpc}$, and most of them have a mass in the range $10^{10}\textup{--}10^{11}~{\rm M_\odot}$. Thus our simulated cluster system well represents those studied in~\cite{Meneghetti:2020yif}.  

\begin{figure*}[htbp]
  \centering
  \includegraphics[scale=0.4]{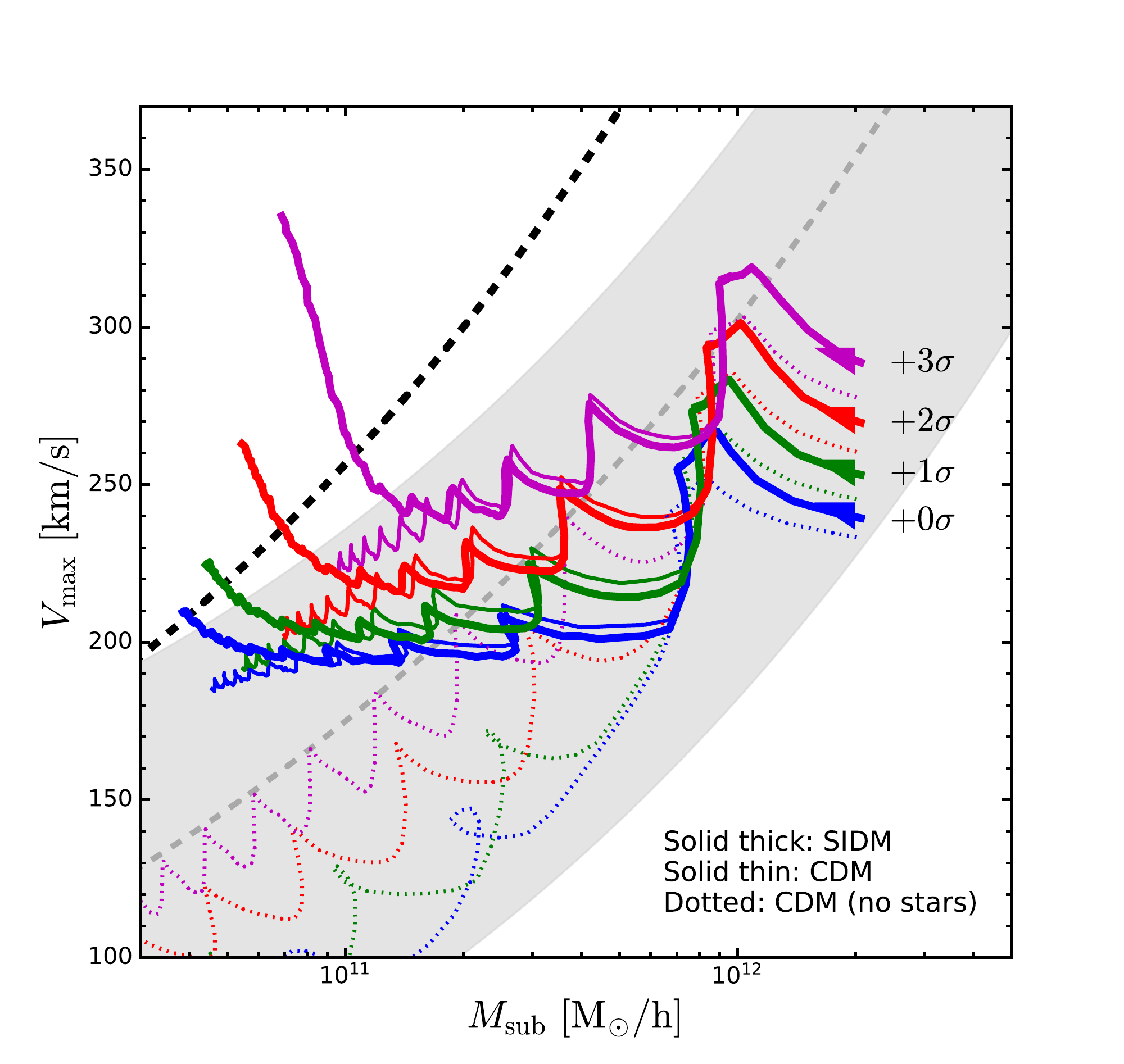}\;\;\;\;\;
  \includegraphics[scale=0.4]{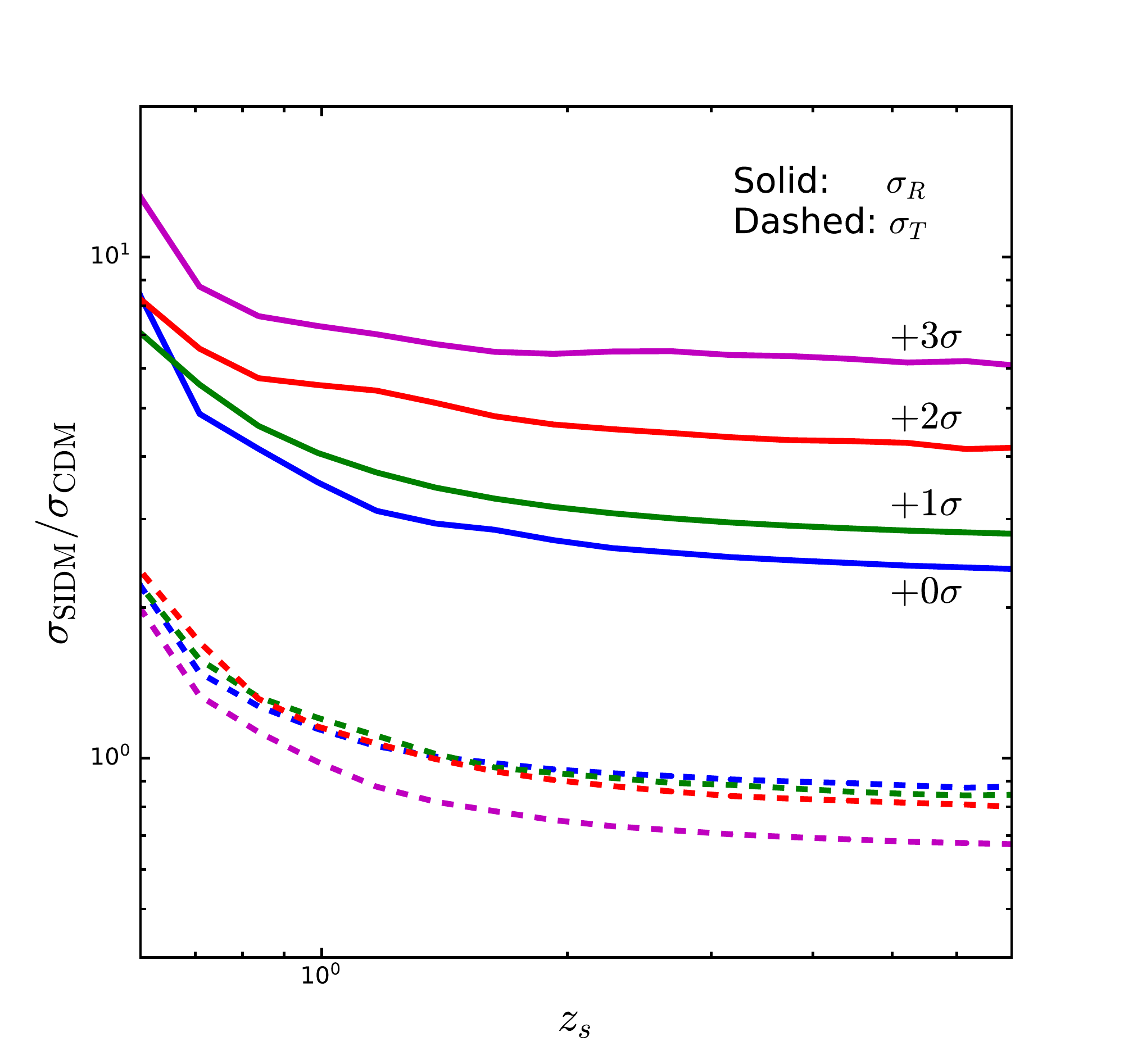}
\caption{\label{fig:sim} \emph{Left:} Evolution of $V_{\rm max}\textup{--}M_{\rm sub}$ for the simulated SIDM (solid thick) and CDM (solid thin) benchmarks with different initial subhalo concentrations, denoted using deviations from the cosmological median at $z=0$, {i.e.}, $+3\sigma$ (magenta), $+2\sigma$ (red), $+1\sigma$ (green) and $+0\sigma$ (blue). The arrows denote the direction of the evolution and the final snapshot is at $t=6~{\rm Gyr}$. CDM simulations without including the star component are also shown (dotted). For comparison, the average relation from strong lensing observations (black dashed)~\cite{Meneghetti:2020yif}, the range (gray band) and best-fit model relation (gray dashed) from their CDM simulations are displayed.~\emph{Right:} SIDM tangential (dashed) and radial (solid) GGSL cross sections vs source redshift, normalized to their corresponding CDM counterparts for $t=6~{\rm Gyr}$. The color scheme is the same as the one in the left panel. The simulated CDM substructures without stars have a low surface density and their lensing effect is negligible. 
 }
\end{figure*}

For the subhalos, we use an NFW profile to model their initial dark matter distribution. We fix the initial virial halo mass to be $M_{200}=3\times10^{12}~{\rm M_\odot}$, and choose four benchmark values for the concentration, i.e., $c_{200}=7.49$, $9.65$, $12.4$ and $16.0$, corresponding to $0\sigma$, $1\sigma$, $2\sigma$ and $3\sigma$ higher than the cosmological median at $z=0$~\cite{Dutton:2014xda}, respectively. For each of the benchmarks, we convert their ($M_{200}$, $c_{200}$) to ($\rho_s$, $r_s$) to specify the initial NFW density profile in our simulations. Note for a given set of $\rho_s$ and $r_s$, the interpretation of $c_{200}$ depends on the redshift. Consider $z=2$, at which infall is expected to occur, the concentration of the benchmarks, from low to high, is $-1.5\sigma$, $-0.39\sigma$, $+0.73\sigma$, and $+1.8\sigma$ away from the median ($z=2$)~\cite{Dutton:2014xda}, which are representative. 

We fix the initial stellar mass as $M_\star=6\times10^{10}~{\rm M_\odot}$, expected from the stellar-to-halo mass relation~\cite{2013MNRAS.428.3121M}, and model its distribution with a truncated singular isothermal profile as in~\cite{Meneghetti:2020yif}, $\rho_\star(r)=\rho_0r^4_{\rm cut}/[r^2(r^2_{\rm cut}+r^2)]$, where $\rho_0$ is the density normalization factor and $r_{\rm cut}$ is the cutoff radius. This is consistent with observations of early-type galaxies~\cite{Bolton:2005nf,Treu:2005aw,Koopmans:2006iu,Gavazzi:2007vw}. We take $r_{\rm cut}=6.23~{\rm kpc}$ following the size-mass relation~\cite{2019MNRAS.485..382C}, and $\rho_0=1.26\times10^7~{\rm M_\odot/kpc^3}$. We use live particles for the subhalo and stellar components and perform both SIDM and CDM simulations. For the former, we choose $\sigma/m=1~{\rm cm^2/g}$, approximately the lower limit that could explain observations on galactic scales~\cite{Tulin:2017ara}. For comparison, we also perform CDM simulations without including stars. As we will show that the stellar component is important in producing strong lensing observables. 

We use the public~\texttt{GADGET-2} code~\cite{Springel:2005mi,Springel:2000yr}, and extend it with a module modeling dark matter self-interactions~\cite{Huo:2019yhk,Yang:2020iya}, which has been validated in both gravothermal expansion and collapse regimes with results from~\cite{Ren:2018jpt,Essig:2018pzq,Pollack:2014rja,Koda:2011yb}. We use the code \texttt {SpherIC}~\cite{GarrisonKimmel:2013aq} to generate initial conditions for the simulated substructures. The mass of the simulated particle is $10^{6}~{\rm M_{\odot}}$ for both subhalo and stellar components, and the softening length is $0.2~{\rm kpc}$. The resolution is high enough to avoid numerical artifacts concerning disruption of substructures~\cite{vandenBosch:2016hjf,vandenBosch:2017ynq,vandenBosch:2018tyt}. We let the simulated substructures evolve for $6~{\rm Gyr}$ in the tidal field of MACSJ1206.

{\noindent\bf Gravothermal collapse.} The left panel of Fig.~\ref{fig:sim} shows the evolution of the maximum circular velocity $V_{\rm max}$ vs the total substructure mass $M_{\rm sub}$ for the benchmarks with $c_{200}=16.0$ (magenta)  $12.4$ (red) $9.65$ (green) and $7.49$ (blue) from our SIDM (solid thick) and CDM (solid thin) simulations, where we include both subhalo and stellar components as in~\cite{Meneghetti:2020yif}. The arrow on each curve denotes the direction of the evolution. The subhalos lose the majority of their mass after tidal evolution, while the stellar mass is only reduced by an ${\cal O}(1)$ factor. Our final total stellar and subhalo masses are consistent with those of cluster substructures from the Illustris simulations~\cite{Niemiec:2018vao,Vogelsberger:2014dza}. For comparison, our CDM simulations without including stars are shown (dotted). 

The maximum circular velocities of the CDM substructures decrease continuously, aside from oscillatory features due to tidal interactions. For those with stars, the final $V_{\rm max}$ values are close to the high end of the range predicted in cosmological hydrodynamical simulations~\cite{Meneghetti:2020yif} (gray band) and the average value from the strong lensing observations (black dashed). We also see that the $V_{\rm max}$ values predicted in our CDM simulations without stars are still within the gray band. It implies that a large population of simulated substructures with AGN feedback in~\cite{Meneghetti:2020yif} has diffuse baryon distributions and high dark matter fractions. For our simulated substructures with stars, the baryon mass fraction is $\sim40\%$ after $6~{\rm Gyr}$ of tidal evolution, which is reasonable compared to that of observed cluster galaxies~\cite{Annunziatella:2017svn}.

The simulated SIDM substructures follow a similar trend for most of the evolution time, but their $V_{\rm max}$ values  spike toward the measured ones at late stages, as gravothermal collapse occurs and their central densities increase. At $t=6~{\rm Gyr}$, all four SIDM benchmarks, even the one with a median concentration ($z=0$), are denser than their CDM counterparts with stars. The collapse occurs earlier if $c_{200}$ is higher, leading to a higher density at $6~{\rm Gyr}$, as its timescale is extremely sensitive to the concentration~\cite{Essig:2018pzq}. We find the presence of stellar particles deepens potential and accelerates gravothermal evolution, as in the isolated case~\cite{Elbert:2016dbb,Sameie:2018chj,Feng:2020kxv}. Without stars, the collapse would not occur within $6~{\rm Gyr}$ unless $c_{200}$ is $6\sigma$ higher than the median, and a subhalo with median $c_{200}$ would be nearly destroyed~\cite{Yang:2020iya}. In the cluster environment, tidal stripping could also speed up the onset of gravothermal collapse~\cite{Nishikawa:2019lsc,Sameie:2019zfo,Kahlhoefer:2019oyt,Correa:2020qam,Turner:2020vlf}.

{\noindent\bf Strong lensing observables.} To further see implications for strong lensing observations, we compute GGSL cross sections for the simulated substructures. We adopt thin-lens approximation and project the mass distribution of the host cluster and substructure, assumed to be spherical, onto the lens plane, which is perpendicular to the line of sight. The distance between the substructure and the host center is fixed to be $300~{\rm kpc}$. We denote the angular positions as $\vect{\theta}$ and $\vect{\beta}$ on the lens and source planes, respectively. It is convenient to introduce effective lensing potential as~\cite{1992grle.book.....S,Kneib:2012ip}:
\begin{eqnarray}
\Psi(\vect{\theta})   &\equiv& \dfrac{1}{\pi} \int d^2 \theta \ln|\vect{\theta}-\vect{\theta}'| \kappa(\vect{\theta}'), \nonumber
\end{eqnarray}
where $\kappa(\vect{\theta}) = \Sigma(D_L \vect{\theta})/\Sigma_{cr}$ is the scaled projected density and $\Sigma_{cr} ={c^2 D_S}/{4\pi G D_L D_{LS}}$ is the critical density. $D_L$ and $D_S$ are lens and source angular diameter distances, respectively, and $D_{LS}$ the distance between the two. We calculate these quantities in a flat universe with matter energy density $\Omega_m=0.3175$ and $h=0.671$~\cite{Ade:2013zuv}. The lens equation is $\vect{\beta} = \vect{\theta} - {D_{LS}}\hat{\vect{\alpha}}/{D_S}$, where $\hat{\vect{\alpha}}$ is the deflection angle.

\begin{figure}[htbp]
  \centering
  \includegraphics[scale=0.40]{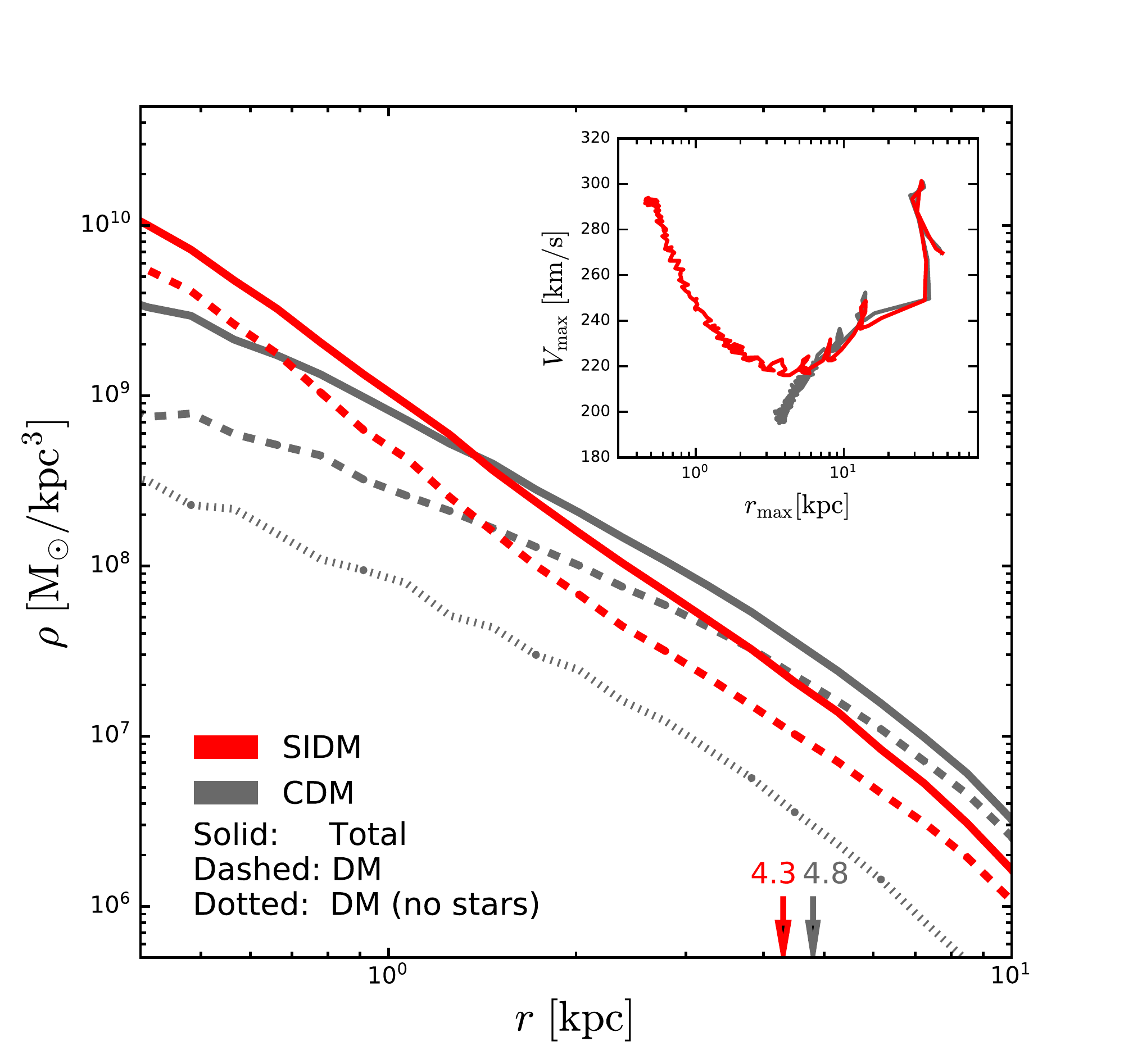}
\caption{\label{fig:prof} Total (solid) and dark matter (dashed) density profiles for the SIDM (red) and CDM (gray) benchmarks with $c_{200}=12.4$ ($+2\sigma$) at $t=6~{\rm Gyr}$. The red and gray arrows denote their Einstein radii, respectively, assuming $z_s=3$. The dark matter density profile from CDM simulations without stars is also shown (dotted); its lensing effect is negligible. \emph{Insert:} Evolution of $V_{\rm max}\textup{--}r_{\rm max}$ for the benchmarks with stars. 
}
\end{figure}

\begin{figure*}[htbp]
  \centering
  \includegraphics[scale=0.4]{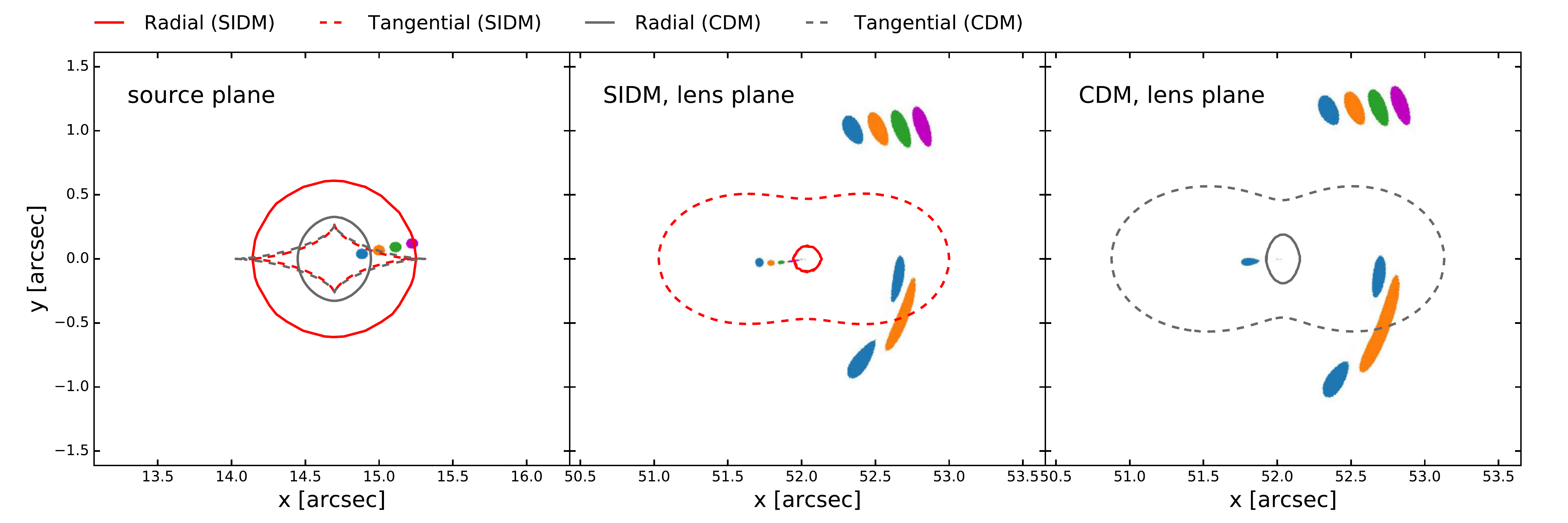}
\caption{\label{fig:images} {\em Left:} Tangential (dashed) and radial (solid) caustics for the simulated SIDM and CDM substructures with $c_{200}=12.4$ ($+2\sigma$), together with four mock sources (circles). The corresponding critical lines and lensed images are shown in the {\em Middle} and {\em Right} panels, respectively, assuming $z_s=3$. For the lens redshift $z_l=0.439$ of MACSJ1206, one arcsec corresponds to $5.76~{\rm kpc}$.}
\end{figure*}

For each simulated substructure, we model its total density profile with a numerical interpolation function and construct its surface density assuming spherical symmetry. We have checked that the substructure slightly deviates from spherical symmetry after tidal evolution, but the deviation is minor and it has negligible effects on the lensing observables. For a  substructure {\it plus} its host cluster, we determine the lensing potential by solving the Poisson equation $\nabla_{\vect{\theta}}^2 \Psi(\vect{\theta}) = 2 \kappa(\vect{\theta})$. We implement the fast Fourier transformation method on a $2000\times 2000$ grid, and use the software~\texttt{Mathematica 12} for numerical computation. 

After obtaining $\Psi$, we calculate the shear matrix as ${\cal A} \equiv {\partial \beta_i}/{\partial \theta_j} = \left(\delta_{ij} - \Psi_{ij} \right)$ and the pseudo-vector shear $\vect{\gamma} = \sqrt{[({\Psi_{11} - \Psi_{22}})/{2}]^2+\Psi_{12}^2}$, where $\Psi_{ij} \equiv {\partial^2 \Psi}/{\partial \theta_i \partial \theta_j}$ and $i,j$ are indices of the two spatial coordinates. The tangential and radial critical lines are contours of $\lambda_t = 1-\kappa-\gamma = 0$ and $\lambda_r = 1-\kappa+\gamma = 0$, respectively. We obtain their corresponding caustic lines by mapping them onto the source plane using the lens equation, and compute tangential and radial GGSL cross sections defined as the area enclosed by the secondary caustic~\cite{Meneghetti:2013sja}. We have further performed convergence tests and confirmed lensing results discussed below are robust.

The right panel of Fig.~\ref{fig:sim} shows ratios of SIDM to CDM tangential (dashed) and radial (solid) GGSL cross sections as a function of the source redshift $z_s$. The tangential cross sections are comparable for both cases and the differences are within order unity. Interestingly, the SIDM substructure has larger a radial cross section than its CDM counterpart, by a factor of $\sim 3\textup{--}7$ for $z_s\gtrsim1$, and the difference increases with the concentration. But this effect is hard to observe as we will discuss later. For the CDM substructures without stars, their surface density is low and the lensing effect is negligible. 

We see the SIDM substructure can be as compact and dense as its CDM counterpart in reproducing small-scale lenses of galaxy clusters. In contrast to the general expectation, the lensing excess reported in~\cite{Meneghetti:2020yif} does not rule out SIDM. Instead, it could be an indicator of gravothermal catastrophe induced by dark matter self-interactions. We also find a compact stellar density profile is necessary in reproducing observed strong lenses in both SIDM and CDM; see also~\cite{Meneghetti:2020yif,Bahe:2021bcs,Robertson:2021} for related discussions.

Although our simulations are based on a particular choice of the cross section, i.e., $\sigma/m= 1~{\rm cm^2/g}$, the lower limit to be relevant for dwarf galaxies, the benchmarks capture all the essential features predicted in SIDM. For example, a larger cross section in the subhalo would further shorten the collapse timescale, resulting in an even high central density. Since there is a strong constraint of $\sigma/m\lesssim0.1~{\rm cm^2/g}$ for cluster halos with masses $\sim10^{15}~{\rm M_\odot}$~\cite{Kaplinghat:2015aga,Sagunski:2020spe,Andrade:2020lqq}, we consider a velocity-dependent SIDM scenario, which can be naturally realized in many particle physics models, see~\cite{Tulin:2017ara}.

{\noindent\bf Density profiles.} We take the benchmark with $c_{200}=12.4$ ($+2\sigma$) and perform a detailed case study. In Fig.~\ref{fig:prof}, we show its dark matter (dashed) and total (solid) density profiles with (red) and without (gray) dark matter self-interactions. After $6~{\rm Gyr}$ of evolution, the collapse leads to an overdense region within $1.3~{\rm kpc}$, and a less dense region $r\gtrsim1.3~{\rm kpc}$, compared to the CDM subhalo. For the CDM substructure  without including stars (dotted), the density is significantly lower. The inset displays the evolution of $V_{\rm max}$ and $r_{\rm max}$ for SIDM and CDM substructures with stars, their $r_{\rm max}$ values decrease overall due to tidal mass loss. The SIDM one becomes further smaller at late stages, as the collapse occurs and the central density increases; see also~\cite{Turner:2020vlf}. 

The left panel of Fig.~\ref{fig:images} shows the corresponding secondary tangential (dashed) and radial (solid) caustics, assuming a source at $z_s=3$ for the SIDM (red) and CDM (gray) benchmarks ($c_{200}=12.4$). For SIDM, the enclosed area of the radial caustic is much larger than the tangential one. For CDM, they are comparable, but both are smaller than the area of the SIDM radial caustic. For a collapsed SIDM substructure, the total density profile is steeper than the isothermal profile $r^{-2}$, and hence its associated radial caustic is larger than the tangential one.

{\noindent\bf Mock lensed images.} The left panel of Fig.~\ref{fig:images} displays four mock sources at four representative locations; the middle (SIDM) and right (CDM) panels show their corresponding lensed images, together with the critical lines mapped from the caustics using the lens equation. The innermost blue source is inside radial {\em and} tangential caustics predicted in SIDM and CDM, and it has four images in both cases~\cite{Blandford:1986,Schneider:2006}. The source in orange sits on the second fold caustic in SIDM while it only crosses the tangential caustic in CDM, thus it has one more image in the former case. Similarly, the sources in green and magenta have one more image in SIDM than in CDM. Our example explicitly illustrates that the collapsed SIDM substructure has a higher capability of producing multiple images. Observationally, the predicted images in the central region are difficult to detect because they are highly demagnified and obscured by bright objects. Thus it remains challenging to differentiate a collapsed SIDM substructure from its CDM counterpart with lensing observations. 

We compute the Einstein radii as $r_E=\theta_E D_L\approx4.3~{\rm  kpc}$ and $4.8~{\rm kpc}$ for the SIDM and CDM benchmarks with stars, respectively, denoted in Fig.~\ref{fig:prof} with arrows, where $r_{E}$ is defined as the radius of the circle with the same area as that enclosed by the critical line~\cite{Meneghetti:2013sja}. It's not surprising that they are comparable. In our setup, the stellar component dominates the inner region and both cases have similar stellar distributions after tidal evolution. In addition, during the collapse process, the SIDM central density increases, but the total mass does not change. As indicated in Fig.~\ref{fig:prof}, $r_{\rm max}\sim r_{E}$ for CDM, while $r_{\rm max}< r_{E}$ for SIDM. Thus the mass distribution induced by gravothermal collapse does not produce an appreciable change in the enclosed mass within $r_E$. 

{\noindent\bf Discussion and Conclusions.} Our simulations assume a compact stellar distribution motivated by observations of early-type galaxies. Hydrodynamical simulations show that for field SIDM halos with masses $\gtrsim10^{12}~{\rm M_\odot}$ stars could dominate the inner region~\cite{Robertson:2017mgj,DiCintio:2017zdz,Despali:2018zpw,Robertson:2020pxj,Sameie:2021ang}. Thus our assumption is well justified. In addition, the SIDM halo structure is more resilient to feedback than its CDM counterpart, because of rapid energy redistributions induced by the self-interactions; see~\cite{Robles:2017, Fitts:2018ycl}. We expect our overall predictions are robust, and it would be interesting to further test them with cosmological simulations. In summary, we have shown that a collapsed SIDM substructure could have a steep density profile, to be consistent with small-scale gravitational lenses observed in galaxy clusters. The features of strong lensing observables predicted in the SIDM scenario could be further tested using existing data~\cite{Talbot:2020arv}, and upcoming observations~\cite{Drlica-Wagner:2019xan}. 

\section*{Acknowledgments}
\vspace{-3mm}

We thank Haipeng An, Seong Chan Park, and Massimo Meneghetti for useful discussions. DY was supported by NSFC under Grant No. 11975134 and the National Key Research and Development Program of China under Grant No.2017YFA0402204. HBY was supported by the U.S. Department of Energy under Grant No. de-sc0008541 and the John Templeton Foundation under Grant ID\# 61884. The opinions expressed in this publication are those of the authors and do not necessarily reflect the views of the John Templeton Foundation.

\bibliographystyle{apsrev}
\bibliography{reference}

\end{document}